\documentclass[conference]{IEEEtran}
\IEEEoverridecommandlockouts
\usepackage{algorithm}
\usepackage{algpseudocode}
\usepackage{amsmath}
\usepackage{amssymb}
\usepackage{cite}
\usepackage{amsmath,amssymb,amsfonts}
\usepackage{subcaption}
\usepackage{graphicx}
\usepackage{anyfontsize}

\usepackage{graphicx}
\usepackage{caption}
\usepackage{subcaption}
\usepackage{float}

\usepackage{microtype}
\usepackage{mathtools}
\usepackage{booktabs} 
\usepackage{textcomp}
\usepackage{xcolor}
\usepackage{placeins}
\usepackage{caption}
\usepackage{amsmath}
\usepackage{float}
\usepackage{balance}
\usepackage{mathrsfs}
\begin{document}
\title{{Efficient and Sustainable Task Offloading in UAV-Assisted MEC Systems via Meta Deep Reinforcement Learning}\\
}
\author{\IEEEauthorblockN{Maryam Farajzadeh Dehkordi}
\IEEEauthorblockA{\textit{Department of Electrical and Computer Engineering} \\
\textit{George Mason University}\\
Fairfax, VA, USA \\
mfarajza@gmu.edu}
\and
\IEEEauthorblockN{ Bijan Jabbari}
\IEEEauthorblockA{\textit{Department of Electrical and Computer Engineering} \\
\textit{George Mason University}\\
Fairfax, VA, USA \\
bjabbari@gmu.edu}
}

\maketitle
\begin{abstract}
Integrated into existing Mobile Edge Computing (MEC) systems, Unmanned Aerial Vehicles (UAVs) serve as a cornerstone in meeting the stringent requirements of future Internet of Things (IoT) networks. The current endeavor studies an MEC system, in which a computationally-empowered UAV, wirelessly linked to a cloud server, is destined for task offloading in uplink transmission of IoT devices. The performance of this system is studied by formulating a resource allocation problem, which aims to maximize the long-term computed task efficiency, while ensuring the stability of task buffers at the IoT devices, UAV and cloud. The problem jointly optimizes the uplink transmit power of IoT devices and their offloading decisions, the trajectory of the UAV and computing power at all transceivers. Regarding the non-convex and stochastic nature of the problem, we devise a multi-step solution approach. Initially, by invoking the fractional programming and Lyapunov theory, we transform the long-term optimization problem into an equivalent per-time-slot form. Subsequently, we recast the reformulated problem as a Markov Decision Process (MDP), which reflects the network dynamics. The MDP model, eventually, serves for training a Meta Twin Delayed Deep Deterministic Policy Gradient (MTD3) agent, in charge of adaptive resource allocation with respect to the MEC system variations derived from the mobility of the UAV and IoT devices. Simulations reveal the dominance of our proposed resource allocation approach over its Deep Reinforcement Learning (DRL)-powered counterparts, increasing computed task efficiency and reducing task buffer lengths.
\end{abstract}
\section{Introduction}
The rapid growth of Internet of Things (IoT) technology has significantly escalated the need for computational power, especially for low-delay applications like autonomous vehicles, where failure is not even an option at scale. However, many IoT devices lack the necessary computing capabilities. Mobile Edge Computing (MEC) systems address this limitation by allowing these devices to offload tasks to nearby, more powerful edge nodes \cite{Wang2023Task}. Unmanned Aerial Vehicles (UAVs) further enhance MEC by offering mobility and flexible deployment. In this context, an efficient resource allocation structure plays a crucial role in managing the demands in a timely approach, which is essential for meeting the stringent requirements of IoT applications.
In this regard, extensive research has examined UAV-assisted MEC systems \cite{Deep2022Zhang, Optimizing2021Zhang, Online2021Hu, Stochastic2019Zhang, Dynamic2022Yang, Poursiami}. Although providing valuable insights, existing work often overlooks the long-term challenges of non-deterministic conditions in these networks. Recent studies address this by exploring UAV-enabled networks through a long-term causality lens, incorporating Lyapunov optimization and then successive convex approximation \cite{Online2021Hu, Stochastic2019Zhang, Dynamic2022Yang}. However, conventional algorithms can be slow, computationally heavy, and less flexible in dynamic settings. To overcome these limitations, advanced reinforcement learning approaches, have been proposed to enhance adaptability and efficiency in real-time UAV operations.
\par In this paper, we propose a three-tier UAV-assisted MEC system with cloud, UAV, and IoT devices tiers managing computational task queues, extending our previous work \cite{dehkordi2024joint}. Unlike prior research focused on short-term metrics, this study emphasizes long-term system performance through a new metric—computed task efficiency, defined as the ratio of time-averaged computed task to transmission latency. To maintain long-term stability, we use fractional programming \cite{dinkelbach1967nonlinear} and Lyapunov optimization \cite{neely2022stochastic}, techniques applied effectively in previous studies \cite{Online2021Hu, Stochastic2019Zhang, Dynamic2022Yang}. However, rather than their complex algorithms, we reformulate the problem as a Markov Decision Process (MDP) and solve it using Deep Reinforcement Learning (DRL) technique that offer an efficient alternative, enabling effective sequential decision-making under uncertainty. Furthermore, to overcome the challenges of limited scalability and poor generalizability associated with deep Q-networks \cite{dehkordi2024joint}, and to enhance accuracy through continuous decision variables, by integrating Meta learning \cite{meta} with Twin Delayed Deep Deterministic Policy Gradient (TD3) \cite{Rahmani2022}, in this paper we introduce Meta Twin Delayed Deep Deterministic Policy Gradient (MTD3) algorithm for UAV-assisted MEC systems. This algorithm optimally balances system utility and stability, adapting effectively to the dynamic movement of UAV. Simulation results show that our resource allocation scheme improves system performance, increasing computed task efficiency and reducing queue lengths.
\section{System Setup and Resource Allocation}\label{sec:System Setup and Resource Allocation}
\begin{figure}
    \centering
    \captionsetup{justification=centering}
    \includegraphics[width=0.7\linewidth]{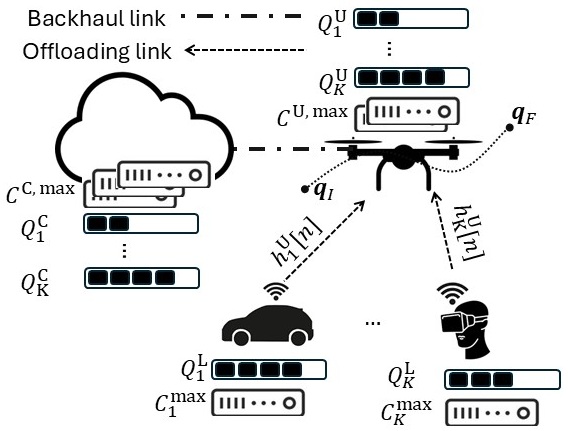}
    \caption{Illustration of a UAV-assisted MEC system}
    \label{fig:system_model}
\end{figure}

As illustrated in Fig. \ref{fig:system_model}, the MEC system comprises a set $\mathcal{K}=\{1, 2, \dots, K\}$ of $K$ IoT devices, each with limited computing power denoted by $C_k^{\text{max}}$, necessitating efficient offloading mechanism for computational tasks. To support this, the MEC network incorporates a cloud infrastructure with high computational capacity denoted by $C^{\text{C, max}}$ capable of computing these tasks. However, the cloud’s considerable distance from the IoT devices introduces communication challenges. To address this, a UAV equipped with computation as well as communication resources is deployed to act as both a relay for re-offloading tasks to the cloud and as an independent computing unit with computational power capacity denoted by $C^{\text{U, max}}$. The system, therefore, follows a three-tier structure: each IoT device computes part of its task locally, offloads another portion to the UAV through via Orthogonal Frequency Division Multiple Access (OFDMA), and queues any remaining tasks. Next, the UAV, in turn, computes part of the offloaded task, re-offloads another portion to the cloud, and retains the rest in its queue. Subsequently, the cloud computes a portion of the re-offloaded tasks and queues any remaining ones. Each device is equipped with a single antenna, and the UAV-to-cloud link operates with sufficient bandwidth.
\subsection{Communication Model}
The UAV’s mission duration \(T\) is divided into \(N\) time slots, each lasting $\frac{T}{N}=\tau$, where each time slot \(n \in \mathcal{N} = \{1, 2, \dots, N\}\). To ensure quasi-static wireless channels within each time slot, we adopt a block fading channel model, in which channel conditions remain constant throughout a slot but may vary independently between successive time slots. Additionally, we assume perfect channel state information, achievable through established estimation methods \cite{Secure2022Lu}. The uplink channels experience both large-scale path loss and small-scale Rician fading with a Rician factor \( K^{\text{R}} \). For the \( k^{\text{th}} \) IoT device in the \( n^{\text{th}} \) time slot with horizontal location at \( [x_k[n],y_k[n]] \), positioned at a distance \( d_k^{\text{U}}[n] \) from the UAV, the distance is given by $d_k^{\text{U}}[n] = \sqrt{(x_k[n] - x^{\text{U}}[n])^2 + (y_k[n] - y^{\text{U}}[n])^2 + H^2}$, where the UAV's location is denoted by \( [x^{\text{U}}[n], y^{\text{U}}[n], H] \). The channel loss is then modeled as $h_k^{\text{U}}[n] \!=\! \sqrt{\frac{\eta_0}{d_k^{\text{U}}[n]^{\theta}}} \left( \sqrt{\frac{K^{\text{R}}}{K^{\text{R}}+1}} \, \rho_k^{\text{LoS}}[n] \!+\! \sqrt{\frac{1}{K^{\text{R}}+1}} \, \rho_k^{\text{NLoS}}[n] \right)$, \cite{3D2019You}, where $\eta_0$ denotes the average channel power gain at the reference distance \(d_0 = 1\) m, and $\theta$ represents the path loss exponent. As such, the UAV receives signals offloaded by all IoT devices at the \(n^{\text{th}}\) time slot, represented as $w[n] = \sum_{k \in \mathcal{K}} h_k^{\text{U}}[n] \sqrt{p_k[n]} u_k[n] + z^{\text{U}}$. Here, \( u_k[n] \) is the unit-power information symbol for the \(k^{\text{th}}\) IoT device at the \(n^{\text{th}}\) time slot, and \( p_k[n] \) denotes the corresponding uplink transmit power. Additionally, \( z^{\text{U}} \sim \mathcal{CN}(0, \sigma_z^2) \) represents the additive white Gaussian noise at the UAV, with zero mean and variance \(\sigma_z^2\). Accordingly, the signal-to-interference-plus-noise ratio for the signal sent by the \(k^{\text{th}}\) IoT device and received by the UAV via OFDMA during this time slot is expressed as $\gamma_k[n] = \frac{\left\lvert h_k^{\text{U}}[n] \right\rvert^2 p_k[n]}{\sum_{\substack{i=1 \\ i \ne k}}^K \left\lvert h_k^{\text{U}}[n] \right\rvert^2 p_i[n] + \sigma_z^2}$. Following the Shannon-Hartley theorem, the maximum achievable task offload rate for the \( k^{\text{th}} \) IoT device at the \( n^{\text{th}} \) time slot is calculated as $R_{k}[n] = B_0 \log_{2}\left( 1 + \gamma_k[n] \right)$, where \( B_0 \) denotes the communication channel bandwidth in Hertz. Then, the transmission latency for offloading the task from the IoT device to the UAV is given by $t_{k}^{\text{U, Comm}}[n] = \frac{x_k^{\text{U}}[n] Q_{k}^{\text{L}}[n]}{R_k[n]}$, where $Q_{k}$ represents the backlog at local tier for the \( k^{\text{th}} \) IoT device at the \( n^{\text{th}} \) time slot and $x_k^{\text{U}}[n]$ denotes the task offloading indicator from the the \( k^{\text{th}} \) IoT device to the UAV at the \( n^{\text{th}}\). Accordingly, the numerator represents the volume of task (in bits) offloaded from the IoT device to the UAV. The transmission latency for re-offloading a task from the UAV to the cloud, denoted by \( t_{k}^{\text{C, Comm}}[n] \), is given by \( t_0 \), a fixed time for establishing a high-speed communication link between the UAV and the cloud, provided by \( x_k^{\text{C}}[n] > 0 \); otherwise, it is \( 0 \). Here, \( x_k^{\text{C}}[n] \) is an indicator variable that determines the portion of the \( k^{\text{th}} \) IoT device’s task at the \( n^{\text{th}} \) time slot is further re-offloaded from the UAV to the cloud. Accordingly, the total transmission latency for the \( k^{\text{th}} \) IoT device's task at the \( n^{\text{th}} \) time slot, denoted as \( t_k^{\text{Comm}}[n] \), is the sum of these transmission latencies calculated as $t_k^{\text{Comm}}[n] = t_{k}^{\text{U, Comm}}[n] + t_{k}^{\text{C, Comm}}[n]$. Then, the system transmission latency at the \( n^{\text{th}} \) time slot is given by $t^{\text{Comm}}[n] = \sum_{k \in \mathcal{K}} t_k^{\text{Comm}}[n]$.
\subsection{Computation Model}
To compute the tasks efficiently, each IoT device computes part of its task locally, offloads another portion to the UAV, and queues any remaining portion. Next, the UAV, in turn, computes part of the offloaded task, re-offloads another portion to the cloud, and retains the rest in its queue. Subsequently, if additional computing is required, the tasks are then re-offloaded to the cloud to complete computation, where the cloud computes a portion of the re-offloaded tasks and queues any remaining ones. Let us define \( c^{\text{L}}_k[n] \), \( c^{\text{U}}_k[n] \), and \( c^{\text{C}}_k[n] \) as the Central Processing Unit (CPU) frequencies (in cycles per second) assigned for computing the \(k^{\text{th}}\) IoT device's task at the \(n^{\text{th}}\) time slot, where computation occurs locally on the IoT device, on the UAV, and on the cloud tiers, respectively. The computation latency for computing the \(k^{\text{th}}\) IoT device’s task at the \(n^{\text{th}}\) time slot on the IoT device, UAV, and cloud are represented by \( t_{k}^{\text{L, Comp}}[n] \), \( t_{k}^{\text{U, Comp}}[n] \), and \( t_{k}^{\text{C, Comp}}[n] \), respectively. Let \( w_k^{\text{L}}[n](1 - x_k^{\text{U}}[n]) Q_k^{\text{L}}[n] \) represent the amount of task (in bits) from the \( k^{\text{th}} \) IoT device’s task that is computed locally during the \( n^{\text{th}} \) time slot, where \( w_k^{\text{L}}[n] \) denotes the fraction of task handled locally, and \( x_k^{\text{U}}[n] \) represents the portion offloaded to the UAV. Then, let $B^{\text{Tot}}[n]$ denote the system computed task at the \( n^{\text{th}} \) time slot given by $B^{\text{Tot}}[n] = \sum_{k\in\mathcal{K}}B^{\text{Tot}}_k[n]$, in which  $B_k^{\text{Tot}}[n] = ~ w_k^{\text{L}}[n](1 - x_k^{\text{U}}[n]) Q_k^{\text{L}}[n] + w_k^{\text{U}}[n](1 - x_k^{\text{C}}[n]) Q_k^{\text{U}}[n] + w_k^{\text{C}}[n] Q_k^{\text{C}}[n]$. Additionally, let \( C \) indicate the number of CPU cycles required to compute each bit of task (in cycles per bit). Accordingly, the local computation latency, denoted by \( t_k^{\text{L, Comp}}[n] \), can be expressed as $t_k^{\text{L, Comp}}[n] = \frac{w_k^{\text{L}}[n](1 - x_k^{\text{U}}[n]) Q_k^{\text{L}}[n]C}{c_k^{\text{L}}[n] }$, where \( c_k^{\text{L}}[n] \) represents the computation power allocated by the \( k^{\text{th}} \) IoT device at the \( n^{\text{th}} \) time slot. In a similar way, the computation latency for the tasks computed by the UAV and the cloud, represented by \( t_{k}^{\text{U, Comp}}[n] \) and \( t_{k}^{\text{C, Comp}}[n] \), respectively, can be determined. Here, it is assumed that the IoT device and the UAV are capable of performing computation and task offloading concurrently. As a result, let \( T_k^{\text{Tot}}[n] \) denote the total task completion latency (in seconds) for the \(k^{\text{th}}\) IoT device’s task during the \(n^{\text{th}}\) time slot, expressed as  $T_k^{\text{Tot}}[n] \!=\! \max \Big\{ t_{k}^{\text{L, Comp}}[n], t_{k}^{\text{U, Comm}}[n] \Big\}+ t_{k}^{\text{C, Comp}}[n] + \!\max \left\{ t_{k}^{\text{U, Comp}}[n], t_{k}^{\text{C, Comm}}[n] \right\}$.
\subsection{Queueing Model}
Each IoT device maintains a dedicated queue to temporarily hold incoming tasks before they are computed. Similarly, both the UAV and cloud server have \( K \) distinct queues, each assigned to a specific IoT device. In this section, we model the queue dynamics and updates at each tier (local, UAV, and cloud).
\par \textbf{Local Queueing Model:} Let \( I_k[n] \) denote the number of task bits arriving at the \( k^{\text{th}} \) IoT devices during the \( n^{\text{th}} \) time slot. The task arrival rate for IoT devices is independent and identically distributed (i.i.d.) and is bounded by a maximum value \( I_k^{\text{max}} \) for each device. Ensuring stability in the queues across all nodes relies on tracking queue backlogs. For each IoT device \( k \) at the \( n^{\text{th}} \) time slot, the backlog at local tier is represented by \( Q_k^{\text{L}}[n] \), initialized at \( Q_k^{\text{L}}[0] = 0 \). As such, a portion of this backlog, offloaded to the UAV, is given by \( x_{k}^{\text{U}}[n] Q_{k}^{\text{L}}[n] \), and accordingly, the portion computed locally by the IoT device is \( w_{k}^{\text{L}}[n](1 - x_{k}^{\text{U}}[n]) Q_{k}^{\text{L}}[n] \). Here, \( w_k^{\text{L}}[n] \) denotes the fraction of task computed by the IoT device itself. Thus, the IoT device queue updates as  $Q_{k}^{\text{L}}[n+1] \!= \left\{ Q_{k}^{\text{L}}[n] \!-\! x_{k}^{\text{U}}[n] Q_{k}^{\text{L}}[n] \!-\! w_{k}^{\text{L}}[n](1 \!-\! x_{k}^{\text{U}}[n]) Q_{k}^{\text{L}}[n] \right\}^{+} + I_k[n]$, where $\{ \cdot \}^{+}$ ensures that queue values remain non-negative.
\par\textbf{UAV Queueing Model:} For the UAV’s queue, which holds tasks offloaded from the \( k^{\text{th}} \) IoT device, the backlog \( Q_k^{\text{U}}[n] \) updates according to  $Q_{k}^{\text{U}}[n+1] = \left\{ Q_{k}^{\text{U}}[n] \!-\! w_k^{\text{U}}[n] (1 - x_k^{\text{C}}[n]) Q_k^{\text{U}}[n] - x_k^{\text{C}}[n] Q_k^{\text{U}}[n] \right\}^{+} + x_{k}^{\text{U}}[n] Q_{k}^{\text{L}}[n]$, where \( w_k^{\text{U}}[n] \) represents the fraction of tasks computed by the UAV, and \( x_k^{\text{C}}[n] \) represents the portion of tasks further re-offloaded to the cloud.
\par \textbf{Cloud Queueing Model:} The cloud queue \( Q_k^{\text{C}}[n] \), storing tasks re-offloaded from the UAV, updates as  $Q_{k}^{\text{C}}[n+1] = \left\{ Q_k^{\text{C}}[n] - w_k^{\text{C}}[n] Q_k^{\text{C}}[n] \right\}^{+} + x_k^{\text{C}}[n] Q_k^{\text{U}}[n]$, where \( w_k^{\text{C}}[n] \) represents the fraction of task computed by the cloud.
Queue stability for each IoT device at local tier is ensured by meeting the condition  $\bar{Q}_k^{\text{L}} = \lim_{N \rightarrow \infty } \frac{1}{N}{\sum_{n \in \mathcal{N}} \left\{ Q_k^{\text{L}}[n] \right\}}  < \infty$, with similar stability conditions for the UAV and cloud queues tiers.
\section{Problem Formulation}\label{subsec: Problem Formulation}
To optimize the balance between the long-term time averaged of system computed task and that of system transmission latency, we define the long-term computed task efficiency as  $J = \frac{\lim\limits_{N \to \infty} \frac{1}{N} \sum\limits_{n \in \mathcal{N}} B^{\text{Tot}}[n]}{\lim\limits_{N \to \infty} \frac{1}{N} \sum\limits_{n \in \mathcal{N}} t^{\text{Comm}}[n]}$.
To optimize \( J \), we adjust system parameters as transmit power matrix, denoted by \( \mathbf{P} = [p_k[n]]\);
offloading decision matrix, represented as \( \mathbf{X} = [\mathbf{x}_k[n]] \), where \( \mathbf{x}_k[n] = [x_k^{\text{U}}[n], x_k^{\text{C}}[n]] \); computing decision matrix, given by \( \mathbf{W} = [\mathbf{w}_k[n]] \) with \( \mathbf{w}_k[n] = [w_k^{\text{L}}[n], w_k^{\text{U}}[n], w_k^{\text{C}}[n]] \); computing power allocation matrix, denoted by \( \mathbf{C} = [\mathbf{c}_k[n]] \), where \( \mathbf{c}_k[n] = [c_k^{\text{L}}[n], c_k^{\text{U}}[n], c_k^{\text{C}}[n]]\); and UAV trajectory vector, represented as \( \mathbf{\hat{q}}[n] = [x^{\text{U}}[n], y^{\text{U}}[n]] \). The problem is then formulated as follows:
\begin{align}
\mathscr{P}_{1}: &\underset{\textbf{P, X, W, C, $\mathbf{\hat{q}}$}}{\max}\quad {J} \nonumber\\
s.t. \quad 
&\text{(\ref{eq:main}a)}\quad 0 \leq p_{k}[n] \leq {P}^{\text{max}}, \quad\forall k \in \mathcal{K}, \; n \in \mathcal{N}, \nonumber\\
&\text{(\ref{eq:main}b)}\quad 0 \leq x_{k}^{\textrm{i}}[n] \leq 1, \quad \forall \textrm{i} \in \{\textrm{U}, \textrm{C}\}, k \in \mathcal{K}, n \in \mathcal{N},\nonumber\\
&\text{(\ref{eq:main}c)}\quad 0 \leq w_{k}^{\textrm{i}}[n] \leq 1, \quad \forall \textrm{i} \in \{\textrm{L}, \textrm{U}, \textrm{C}\}, k \in \mathcal{K}, n \in \mathcal{N}, \nonumber\\
&\text{(\ref{eq:main}d)}\quad 0 \leq c_{k}^{\text{L}}[n] \leq {C}_{k}^{\text{max}}, \quad  \forall k \in \mathcal{K}, \forall n \in \mathcal{N}, \nonumber\\
&\text{(\ref{eq:main}e)}\quad 0 \leq \sum\nolimits_{k \in \mathcal{K}}c_{k}^{\text{U}}[n] \leq {C}^{\text{U, max}}, \quad \forall n \in \mathcal{N},\nonumber\\
&\text{(\ref{eq:main}f)}\quad 0 \leq \sum\nolimits_{k \in \mathcal{K}}c_{k}^{\text{C}}[n] \leq {C}^{\text{C, max}},  \quad \forall n \in \mathcal{N},\nonumber\\
&\text{(\ref{eq:main}g)}\quad 0 \leq \frac{\ \left\|\mathbf{\hat{q}}[n+1] - \mathbf{\hat{q}}[n]\right\|}{\tau} \leq V^{\text{max}} ,\quad  \forall n \in \mathcal{N},    \nonumber\\
&\text{(\ref{eq:main}h)}\quad \mathbf{\hat{q}}[1] = \mathbf{q}_I, \ \mathbf{\hat{q}}[N] = \mathbf{q}_F,\nonumber\\
&\text{(\ref{eq:main}i)}\quad T_k^{\text{Tot}}[n] \leq \tau, \quad\forall k \in \mathcal{K}, \; n \in \mathcal{N}, \nonumber\\
&\text{(\ref{eq:main}j)}\quad \bar{Q}_k^{\textrm{i}} < \infty, \quad \forall \textrm{i} \in \{\textrm{L}, \textrm{U}, \textrm{C}\}, k \in \mathcal{K},
\label{eq:main}
\end{align}

Here, $\text{(\ref{eq:main}a)}$ ensures IoT devices transmit power budget at each time slot. $\text{(\ref{eq:main}b)}$ and $\text{(\ref{eq:main}c)}$, specify partial offloading and computing indicators at each time slot, respectively. As well, $\text{(\ref{eq:main}d)}$, $\text{(\ref{eq:main}e)}$ and $\text{(\ref{eq:main}f)}$, respectively assure the computing capacity of $k^{\text{th}}$ IoT device, the UAV, and the cloud, up to $\text{C}_{k}^{\text{max}}$, $\text{C}^{\text{U, max}}$ and $\text{C}^{\text{C, max}}$, at each time slot. In $\text{(\ref{eq:main}g)}$, we have limited the UAV maximum velocity to $V^{\text{max}}$ at each time slot. Also, $\text{(\ref{eq:main}h)}$ mandates that the UAV returns to its starting point, with $\mathbf{q}_I$ and $\mathbf{q}F$ as the UAV’s starting and ending locations at each time slot. Constraint $\text{(\ref{eq:main}i)}$ imposes a maximum allowable delay, denoted by $\tau$, for computing each task. Finally, constraint $\text{(\ref{eq:main}j)}$ guarantees the stability of task queues for IoT devices, the UAV, and the cloud tiers.
\section{per-time-slot Transformations}
Due to the reliance on time-averaged values of the computeded task and transmission latency in the objective function and queue length across the local, UAV, and cloud tiers in the queue stability constraint, the problem becomes hard to solve. The reason is that these values depend on information that is unavailable at the current time and cannot be reliably approximated because of the high dynamicity of the environment. To overcome this, we need to relax the long-term considerations before proceeding with a solution. Subsequently, in this section, we use an tractable approximation alongside the fractional-programming theory to relax long-term consideration of the system utility and Lyapunov function to relax long-term constraint of queue stability at each tier. Subsequently, the optimization problem $\mathcal{P}_1$ is converted into a sequence of per-time-slot subproblem.
\subsection{Objective Function Transformation}
To make the objective function \( J \) tractable at the current time, we express a tractable system utility \( J[n] \) at the \( n^{\text{th}} \) time slot as $J[n] = \frac{\sum_{m=0}^{n-1} B^{\text{Tot}}[m]}{\sum_{m=0}^{n-1} t^{\text{Comm}}[m]}$. Note that future information are not needed anymore to calculate this value. Subsequently, leveraging fractional programming \cite{dinkelbach1967nonlinear}, the non-fractional computed task efficiency \( W[n] \) is then given by $W[n] = B^{\text{Tot}}[n] - J[n] t^{\text{Comm}}[n]$ that can well approximate the system utility $J$.
\subsection{Time-Averaged Constraint Transformation}
Similarly, constraint \( \text{(\ref{eq:main}j)} \) in $\mathscr{P}{1}$ is represented in a time-averaged form. To relax this long-term consideration, a lower value of the Lyapunov function, denoted by \( \mathbf{\Phi} [n] \), can lead to reduced backlog and decreased queue congestion, thereby enhancing the overall stability of the network \cite{neely2022stochastic}. Accordingly, Lyapunov function given by  \(L \left( \mathbf{\Phi} [n] \right) = \frac{1}{2} \sum_{k\in\mathcal{K}} \left( \left(Q_k^{\text{L}}[n]\right)^2 + \left(Q_k^{\text{U}}[n]\right)^2 + \left(Q_k^{\text{C}}[n]\right)^2 \right)\), has its one-step conditional drift upper-bounded as $\Delta L \left( \mathbf{\Phi} [n] \right) \leq  A[n]+ B$, where $A[n]=\sum_{k=1}^{K}A_k[n]$, and $A_k[n]$ is given as  $A_k[n]=  x_k^{\text{U}}[n] Q_k^{\text{L}}[n]Q_k^{\text{U}}[n] 
+ x_k^{\text{C}}[n] Q_k^{\text{U}}[n] Q_k^{\text{C}}[n]  
- \left( w_k^{\text{L}}[n](1 - x_k^{\text{U}}[n]) + x_k^{\text{U}}[n] \right) \left(Q_k^{\text{L}}[n]\right)^2 
- w_k^{\text{C}}[n] \left(Q_k^{\text{C}}[n]\right)^2 
- \left( w_k^{\text{U}}[n](1 - x_k^{\text{C}}[n]) + x_k^{\text{C}}[n] \right) \left(Q_k^{\text{U}}[n]\right)^2$, and $B$ denotes a finite constant value. Subsequently, instead of minimizing Lyapunov drift directly, we can minimize its upper bound and the near-optimal system stability can be achieved.
\subsection{Optimization Problem Transformation}
Considering the affect of queue stability into the non-fractional form of the objective function, we convert the original problem $\mathscr{P}{1}$ into a series of per-time-slot optimization problem $\mathscr{P}{2}$, where $V$ represents a non-negative weight constant, establishing a trade-off among the queue stability and the system utility, formulated as:
 \begin{align}
\mathscr{P}_{2}:&\max_{\mathbf{P}, \mathbf{X}, \mathbf{W}, \mathbf{C}, \mathbf{\hat{q}}}-A[n]+v\displaystyle{W}[n]\nonumber  \\
\text{s.t.} & \quad \text{(\ref{eq:main}a)} - \text{(\ref{eq:main}i)} \quad in \quad \mathscr{P}_{1}, \nonumber
\end{align}

where parameter $v$ is a scaling and trade-off factor, obtained experimentally, that ensures the minimization of $A[n]$ and the maximization of $W[n]$ are balanced in magnitude, preventing one term from dominating the objective function. As observed, long-term considerations of problem $\mathscr{P}{1}$ are relaxed in problem $\mathscr{P}{2}$. However, the system transmission latency in the objective function is fractional with respect to optimization parameters such as transmit power, computing portion and offloading indicator, rendering the problem non-convex. While a brute-force exhaustive search could theoretically find a globally optimal solution, it is impractical due to the complexity and scalability of the problem, even for moderately sized systems. Moreover, traditional approaches based on convex optimization theory \cite{Online2021Hu, Stochastic2019Zhang, Dynamic2022Yang} require complex and time-consuming transformations to achieve locally optimal solutions. However, given the dynamic nature of the MEC network, influenced by UAV mobility, a real-time solution is essential.
\section {Proposed Solution Approach}
To address the complex optimization problem $\mathscr{P}{2}$, DRL is an effective approach as it can jointly optimize multiple variables through real-time system feedback while simultaneously learning system dynamics, without decomposing the problem into subproblems. Accordingly, we begin by modeling the system parameters as a MDP \cite{powell2007approximate}. In our previous work \cite{dehkordi2024joint}, we demonstrated the feasibility of this approach using deep Q-networks (DQN). However, DQN's reliance on discrete action spaces limited its ability to handle the continuous optimization parameters required in this setting. Additionally, its scalability was constrained, as the action space grew exponentially with the number of IoT devices, making it impractical for large-scale deployments. To overcome these limitations, we adopt the MTD3 algorithm, which efficiently manages continuous action spaces while improving adaptability in dynamic environments. Meta-learning in MTD3 enables the agent to adaptively fine-tune its critics based on task-specific feedback, leading to faster convergence and improved generalization across varying environments.
\subsection {MDP Model of the Problem}
Let us consider an edge node as the decision-maker agent, responsible for capturing system features and dynamics while optimizing parameters. In this section, we model our wireless system as an MDP, a framework in which the environment is represented by a set of states influenced by the agent’s actions with each state-action pair yielding an associated reward. By receiving this reward value, the agent learns and adapts its policy to improve system performance. These components, tailored specifically to our problem, are defined as follows.
\par \hspace{-10pt}\textbf{State Space:} The state space, denoted as \( \mathcal{S} \), consists of all possible states the system can occupy at any given time. Here, the state is a minimal and sufficient representation of past information, containing everything necessary to make optimal decisions in the present and future \cite{powell2007approximate}. At the \( n^{\text{th}} \) time slot, the system’s state is given by \( \mathbf{s}[n] \in \mathcal{S} \) and expressed as \( \mathbf{s}[n] = \{ n, s_k[n], \forall k \in \mathcal{K} \} \), where \( s_k[n] \) represents the state of the \( k^{\text{th}} \) IoT device. This device state \( s_k[n] \) is detailed as \( s_k[n] = \big[ Q_k^\text{L}[n], Q_k^\text{U}[n], Q_k^\text{C}[n] \big] \), capturing different queue conditions of the device.
\par \hspace{-10pt}\textbf{Action Space:} The set of actions, represented by \( \mathcal{A} \), covers all feasible actions available for each state in the system. Here, an action is a decision variable that drives the evolution of the system from one state to another, representing the choice made by the decision-maker. During the \( n^{\text{th}} \) time slot, the action chosen is represented by \( \mathbf{a}[n] \in \mathcal{A} \) and takes the form \( \mathbf{a}[n] = \{ a_k[n], \forall k \in \mathcal{K} \} \), where \( a_k[n] \) denotes the action taken by the \( k^{\text{th}} \) IoT device, which is defined as \( a_k[n] = \big[ p_k[n], \mathbf{x}_k[n], \mathbf{w}_k[n], \mathbf{c}_k[n], \hat{\mathbf{q}}[n] \big] \), encompassing various control parameters for system operation.
\par \hspace{-10pt}\textbf{Reward:} The reward function, \( r(\mathbf{s}[n], \mathbf{a}[n]) : \mathcal{S} \times \mathcal{A} \rightarrow \mathbb{R} \), assigns a numerical value that quantifies the immediate benefit or cost associated with taking an action to the combination of system state and action in a given state at a time slot \cite{powell2007approximate}. As such, at the \( n^{\text{th}} \) time slot, the reward for our system is defined by \( r(\mathbf{s}[n], \mathbf{a}[n]) = \sum_{k \in \mathcal{K}} r_k(\textbf{s}[n], \textbf{a}[n]) \), where \( r_k(\textbf{s}[n], \textbf{a}[n]) \) is the reward attributed to the \( k^{\text{th}} \) IoT device given by $r_k(\textbf{s}[n], \textbf{a}[n]) =  \eta_k[n](-A_k[n] + v_1 B_k^{\text{Tot}}[n] - v_2Z_k[n] t_k^{\text{Comm}}[n]) $, where $v_1$ and $v_2$ are obtained experimentally, and $Z_k[n] = \tfrac{\sum_{m=0}^{n-1} B^{\text{Tot}}_k[m]}{\sum_{m=0}^{n-1} t^{\text{Comm}}_k[m]}$. Here, $\eta_k[n]$ represents how much of the available time slot remains after accounting for the total task completion latency of the the \( k^{\text{th}} \) IoT device at \( n^{\text{th}} \) time slot, calculated as $\eta_k[n] = [1-\frac{t_k^{tot}[n]}{\tau}]^+$.
\subsection{MTD3 Approach}\label{sec:MTD3}
To address problem $\mathscr{P}{2}$, we propose utilizing MTD3 as detailed in Algorithm \ref{alg:mtd3} . 

\begin{algorithm}
\caption{MTD3 (Meta Twin Delayed Deep Deterministic Policy Gradient)}
\label{alg:mtd3}
\scriptsize 
\begin{algorithmic}[1]
\State Init meta-policy \( \pi_\phi \), meta-critics \( Q_{\theta_1}, Q_{\theta_2} \), target nets \( \phi'\gets\phi, \theta_j'\gets\theta_j \)
\State Init replay buffers \( \mathcal{B}_i \) for each task \( \mathcal{T}_i \)
\For{each meta-iteration}
    \State Sample tasks \( \{ \mathcal{T}_i \}_{i=1}^I \sim p(\mathcal{T}) \)
    \For{each task \( \mathcal{T}_i \)}
        \State Set \( \phi_i \gets \phi, \theta_{i,j} \gets \theta_j \)
        \For{\( t = 1 \) to \( C_{\text{In}} \)}
            \State Take action \( a \sim \pi_{\phi_i}(s) + \epsilon \), observe \( r, s' \), store in \( \mathcal{B}_i \)
            \State Sample batch of size \( M \) from \( \mathcal{B}_i \)
            \State \( a' \gets \pi_{\phi'}(s') + \epsilon \), \( y \gets r + \gamma \min_j Q_{\theta_j'}(s', a') \)
            \State Update critics: \( \theta_{i,j} \gets \arg\min \sum (y - Q_{\theta_{i,j}}(s,a))^2 \)
            \If{\( t \mod d = 0 \)}
                \State \( \phi_i \gets \phi_i + \alpha \nabla_{\phi_i} J(\phi_i) \)
                \State Update targets: \( \theta_j' \gets \tau \theta_{i,j} + (1 - \tau)\theta_j', \phi' \gets \tau \phi_i + (1 - \tau)\phi' \)
            \EndIf
        \EndFor
        \State Compute \( L_{\text{meta},i} = \frac{1}{M} \sum_j \sum (y - Q_{\theta_{i,j}}(s, a))^2 \)
    \EndFor
    \State Meta-update: \( \phi \gets \phi - \alpha \nabla_\phi \sum_i L_{\text{meta},i} \)
    \State \( \theta_j' \gets \tau \theta_{i,j} + (1 - \tau)\theta_j' \)
\EndFor
\end{algorithmic}
\end{algorithm}

The per-iteration time complexity of the MTD3 algorithm is given by  \(\mathcal{O}(C_{\text{In}} M ((|\mathcal{S}| + |\mathcal{A}|) W + (L-1) W^2 + W |\mathcal{A}|))\), where \(C_{\text{In}}\) denotes the number of inner loop steps per task, \(M\) is the mini-batch size, \(|\mathcal{S}|\) and \(|\mathcal{A}|\) are the dimensions of the state and action spaces, \(W\) is the number of neurons per hidden layer, and \(L\) is the number of hidden layers. In our system, where the state and action space dimensions scale linearly with the number of devices \(K\) (i.e., \(|\mathcal{S}| \propto K\) and \(|\mathcal{A}| \propto K\)), an increase in the number of IoT devices leads to a linear growth in \(|\mathcal{S}|\) and \(|\mathcal{A}|\). This linear growth directly results in a proportional increase in the overall computational complexity, significantly affecting the algorithm’s scalability in large-scale IoT deployments.

\section{Numerical Results}
In this section, we illustrate the effectiveness of the proposed resource allocation mechanism by comparing the the system performance against two benchmark schemes Deep Deterministic Policy Gradient (DDPG) \cite{Deep2022Zhang} and Twin Delayed Deep Deterministic Policy Gradient (TD3) \cite{Rahmani2022}.
\begin{figure*}[ht]
\centering
\begin{subfigure}[t]{0.3\textwidth}
    \centering
    \includegraphics[width=\linewidth]{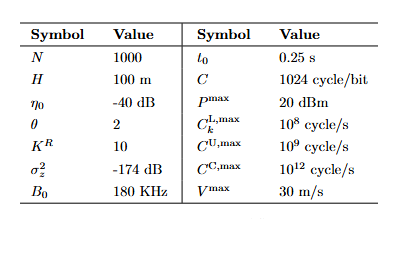}
    \captionsetup{justification=centering}
    \caption{Simulation parameters table}
    \label{table:sim_par}
\end{subfigure}
\hfill
\begin{subfigure}[t]{0.31\textwidth}
    \centering
    \includegraphics[width=\linewidth]{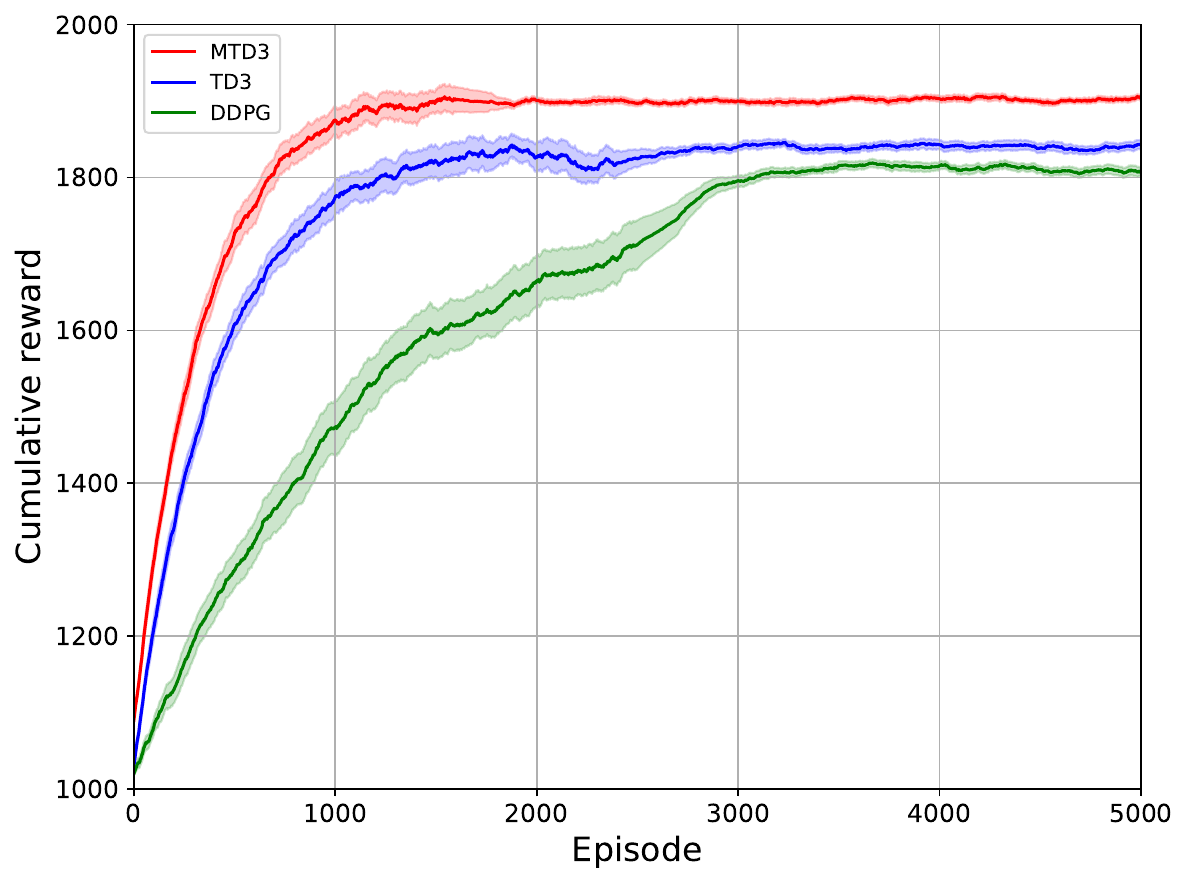}
    \captionsetup{justification=centering}
    \caption{Convergence behavior comparison}
    \label{Fig:Convergence}
\end{subfigure}
\hfill
\begin{subfigure}[t]{0.3\textwidth}
    \centering
    \includegraphics[width=\linewidth]{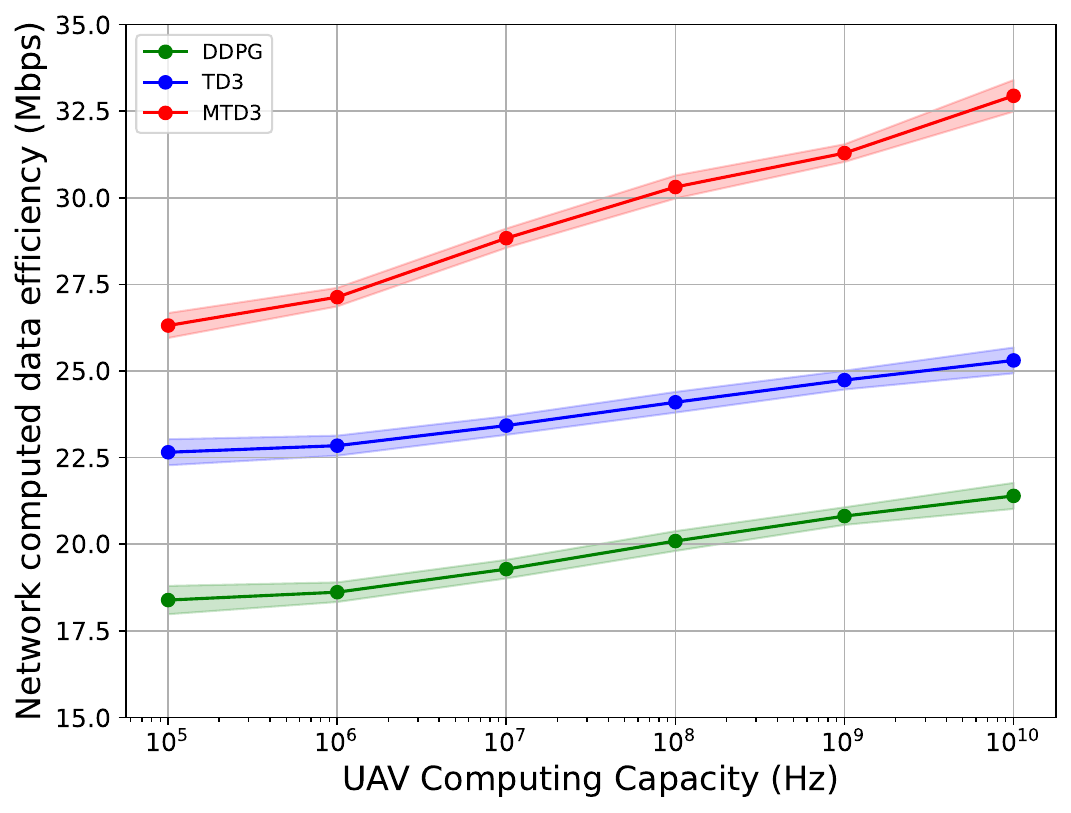}
    \captionsetup{justification=centering}
    \caption{Computed task efficiency}
    \label{fig:PDE_vs_CUmax}
\end{subfigure}
\begin{subfigure}[t]{0.3\textwidth}
    \centering
    \includegraphics[width=\linewidth]{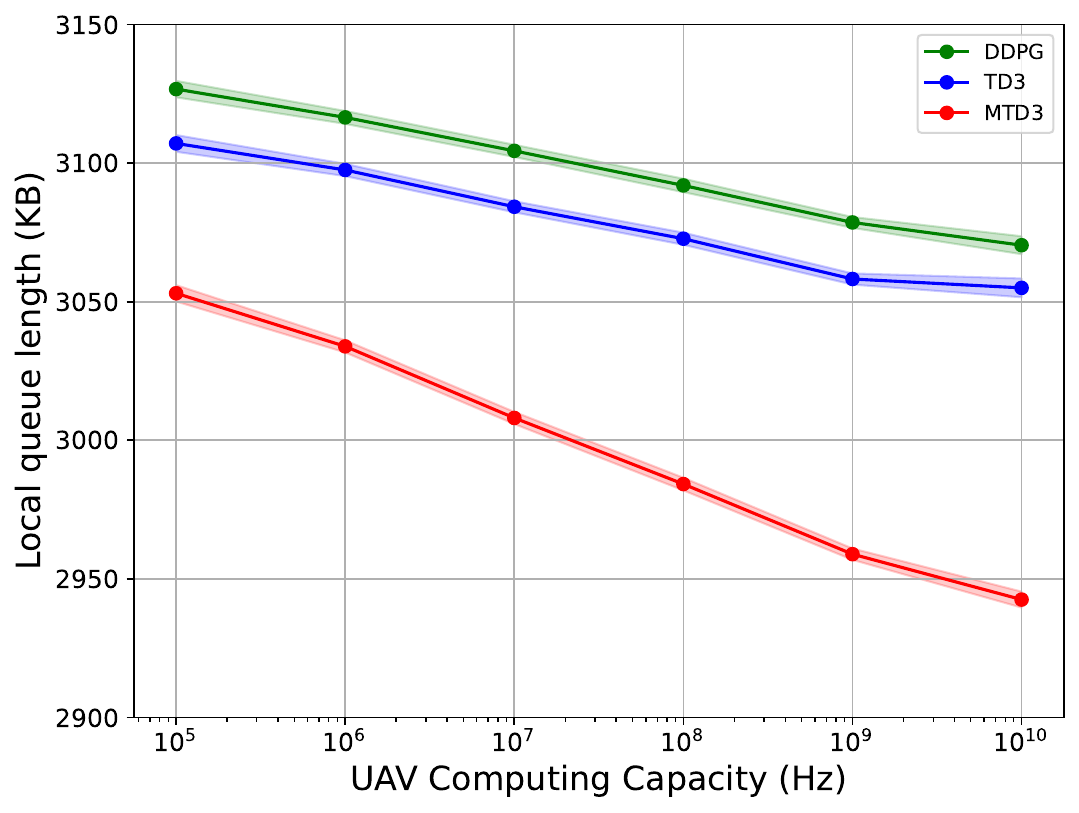}
    \captionsetup{justification=centering}
    \caption{Length of IoT queues}
    \label{fig:local_vs_CUmax}
\end{subfigure}
\hfill
\begin{subfigure}[t]{0.3\textwidth}
    \centering
    \includegraphics[width=\linewidth]{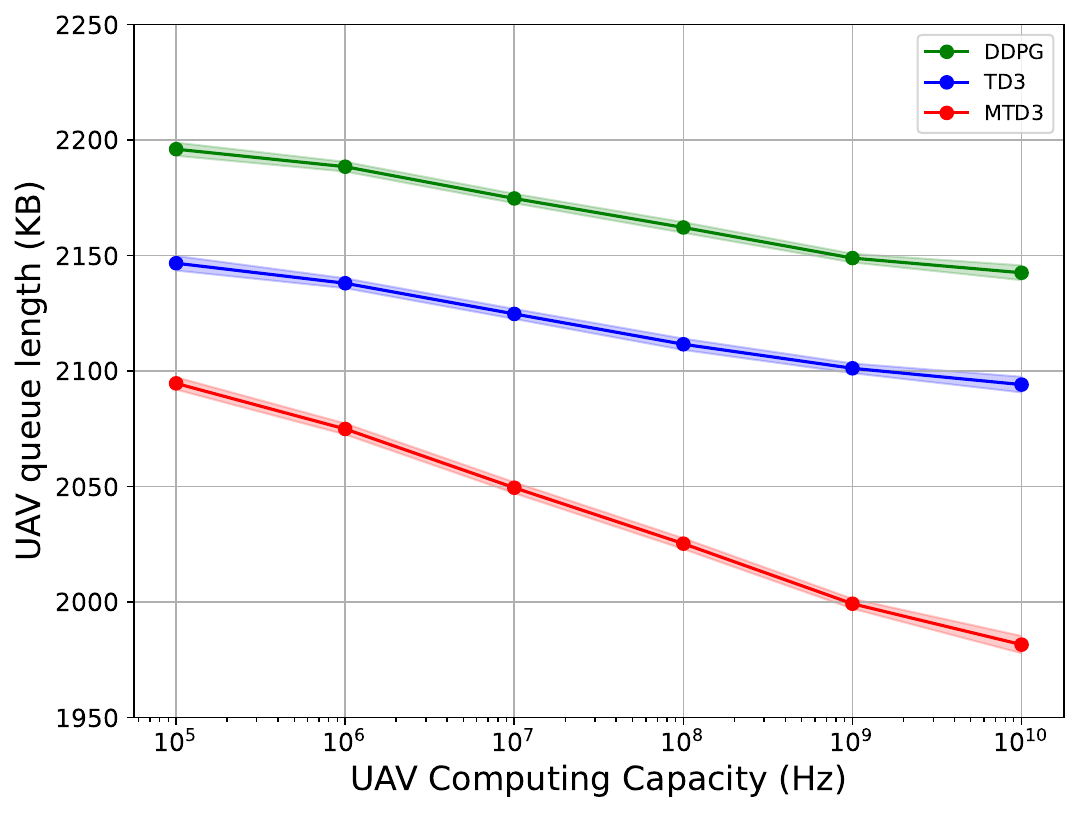}
    \captionsetup{justification=centering}
    \caption{Length of UAV queues}
    \label{fig:uav_vs_CUmax}
\end{subfigure}
\hfill
\begin{subfigure}[t]{0.3\textwidth}
    \centering
    \includegraphics[width=\linewidth]{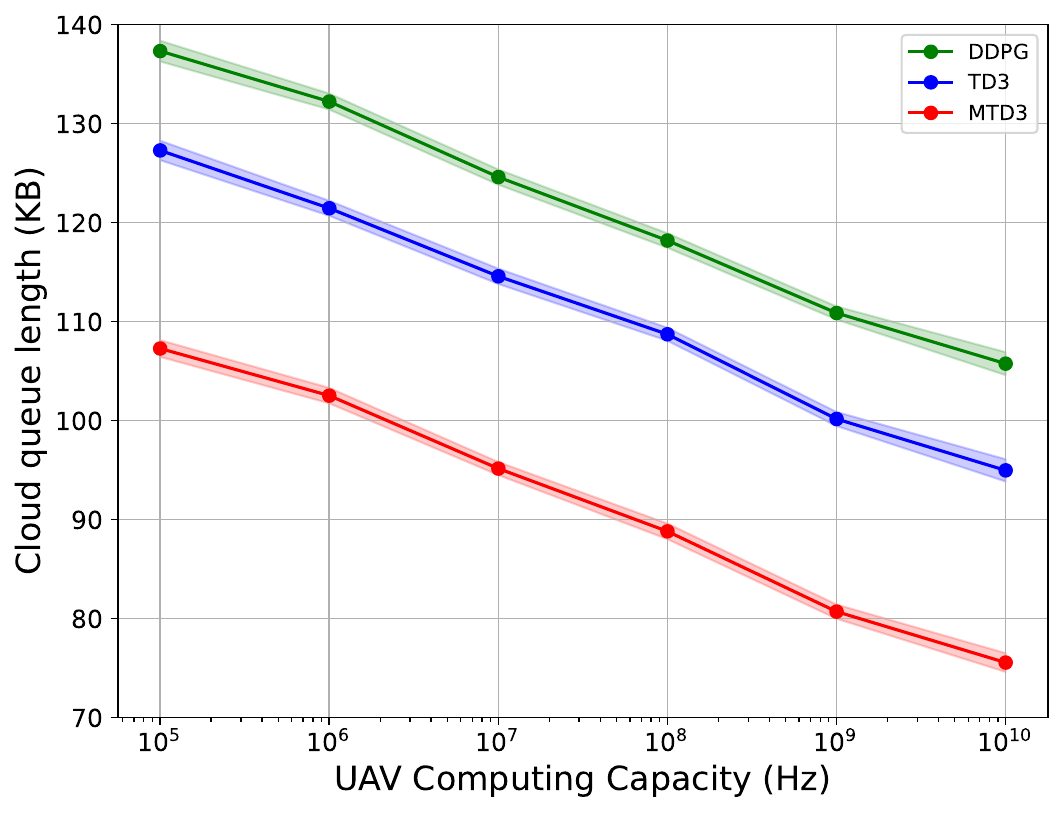}
    \captionsetup{justification=centering}
    \caption{Length of cloud queues}
    \label{fig:cloud_vs_CUmax}
\end{subfigure}
\captionsetup{justification=centering}
\caption{Simulation parameters table, convergence behaviour and performance evaluation comparison of MTD3, TD3, and DDPG across UAV computing capacities.}
\label{fig:performance_comparison}
\end{figure*}
In the simulation, the UAV follows a trajectory that starts and ends at the center of a \(1000 \times 1000 \, \text{m}^2\) coverage area, within which five IoT devices are randomly distributed. Each time slot, denoted by \(\tau\), lasts for 1 second, and the incoming task for each IoT device during each time slot is randomly generated between 0 and 50 MB. The remaining simulation variables are detailed in Table \ref{table:sim_par}, unless stated otherwise. To thoroughly assess the performance of the proposed resource allocation scheme, each scenario was simulated in 50 parallel environments, each for 1000 realizations.

The convergence behavior and performance of MTD3 are compared to baseline algorithms (DDPG and TD3) across various metrics in Figs.~\ref{Fig:Convergence}-\ref{fig:cloud_vs_CUmax}. In Fig.~\ref{Fig:Convergence}, MTD3 converges faster and achieves higher rewards than DDPG and TD3, highlighting its improved sample efficiency and stability during training. This performance boost is primarily due to the integration of meta-learning techniques, which enable MTD3 to adaptively fine-tune its critic networks based on task-specific learning dynamics. The early-stage learning curve of MTD3 shows a steep upward trend, indicating that it quickly adapts to the environment dynamics, whereas TD3 and DDPG exhibit slower, more fluctuating progress. The meta-learning component allows the agent to generalize better across varied state-action distributions, leading to more informed policy updates and accelerated convergence.

Fig.~\ref{fig:PDE_vs_CUmax} illustrates the effect of UAV computing capacity on computed task efficiency. As the UAV’s capacity increases, computed task efficiency (in MBps) also rises, with MTD3 outperforming DDPG and TD3 by 53.40\% and 29.28\%, respectively.  

Fig.s~\ref{fig:local_vs_CUmax}, \ref{fig:uav_vs_CUmax}, and \ref{fig:cloud_vs_CUmax} depict how increasing UAV computing capacity influences queue lengths at the IoT, UAV, and cloud tiers. As UAV computational resources grow, the overall queue lengths decrease, as more tasks are processed locally at the UAV, reducing the workload burden on other tiers.  

In Fig.~\ref{fig:local_vs_CUmax}, higher UAV computational power enables a greater portion of tasks to be offloaded, significantly reducing local queue congestion. MTD3 demonstrates superior performance by reducing local queue lengths by 3.69\% and 4.32\% compared to TD3 and DDPG, respectively. Similarly, Fig.~\ref{fig:uav_vs_CUmax} highlights that as the UAV’s processing capability improves, it handles a larger share of tasks, resulting in shorter UAV queue lengths. MTD3 again outperforms TD3 and DDPG, reducing UAV queue lengths by 5.39\% and 7.58\%, respectively.  

Fig.~\ref{fig:cloud_vs_CUmax} further shows that a more capable UAV relies less on cloud offloading, thereby shortening cloud queue lengths. Here, MTD3 demonstrates up to 28.96\% improvement compared to DDPG and up to 21.37\% improvement compared to TD3. Among the three approaches, DDPG results in the longest queues, indicating relatively inefficient workload distribution.  

\section{Conclusions}
This study has examined the joint optimization of computed task and transmission latency, while ensuring queue stability in a UAV-assisted MEC framework from a long-term perspective. This optimization was structured in three steps relying on Lyapunov theory, MDP reformulation and MTD3. The proposed MTD3 optimization outperformed DDPG and TD3 by up to 53.40\% and 29.28\% in computed task efficiency, respectively. MTD3 also reduces queue lengths by 3.69\%, 5.39\%, and 21.37\% compared to TD3, and by 4.32\%, 7.58\%, and 28.96\% compared to DDPG at local, UAV, and cloud tiers, respectively.
\balance
\bibliographystyle{IEEEtran}

\begin{thebibliography}{15}

\bibitem{Wang2023Task}
K. Wang, J. Jin, Y. Yang, T. Zhang, A. Nallanathan, C. Tellambura, and B. Jabbari, ``Task offloading with multi-tier computing resources in next generation wireless networks,'' \emph{IEEE J. Sel. Areas Commun.}, vol. 41, no. 2, pp. 306--319, Feb. 2023.

\bibitem{Deep2022Zhang}
L. Zhang, B. Jabbari, and N. Ansari, ``Deep reinforcement learning driven {UAV}-assisted edge computing,'' \emph{IEEE Internet Things J.}, vol. 9, no. 24, pp. 25449--25459, Aug. 2022.

\bibitem{Optimizing2021Zhang}
L. Zhang and N. Ansari, ``Optimizing the operation cost for {UAV}-aided mobile edge computing,'' \emph{IEEE Trans. Veh. Technol.}, vol. 70, no. 6, pp. 6085--6093, Jun. 2021.

\bibitem{Online2021Hu}
H. Hu, X. Zhou, Q. Wang, and R. Q. Hu, ``Online computation offloading and trajectory scheduling for {UAV}-enabled wireless powered mobile edge computing,'' \emph{China Commun.}, vol. 19, no. 4, pp. 257--273, Apr. 2022.

\bibitem{Stochastic2019Zhang}
J. Zhang, L. Zhou, Q. Tang, E. C.-H. Ngai, X. Hu, H. Zhao, and J. Wei, ``Stochastic computation offloading and trajectory scheduling for {UAV}-assisted mobile edge computing,'' \emph{IEEE Internet Things J.}, vol. 6, no. 2, pp. 3688--3699, Apr. 2019.

\bibitem{Dynamic2022Yang}
Z. Yang, S. Bi, and Y. A. Zhang, ``Dynamic offloading and trajectory control for {UAV}-enabled mobile edge computing system with energy harvesting devices,'' \emph{IEEE Trans. Wireless Commun.}, vol. 21, no. 12, pp. 10515--10528, Dec. 2022.

\bibitem{Poursiami}
H. Poursiami and B. Jabbari, ``On multi-task learning for energy efficient task offloading in multi-{UAV} assisted edge computing,'' in \emph{Proc. IEEE Wireless Commun. Netw. Conf. (WCNC)}, Dubai, United Arab Emirates, 2024, pp. 1--6.

\bibitem{dehkordi2024joint}
M. F. Dehkordi and B. Jabbari, ``Joint long-term processed task and communication delay optimization in {UAV}-assisted {MEC} systems using {DQN},'' in \emph{Proc. IEEE Mil. Commun. Conf. (MILCOM)}, Washington, DC, USA, 2024, pp. 288--293.

\bibitem{dinkelbach1967nonlinear}
W. Dinkelbach, ``On nonlinear fractional programming,'' \emph{Manage. Sci.}, vol. 13, no. 7, pp. 492--498, 1967.

\bibitem{neely2022stochastic}
M. J. Neely, ``Stochastic network optimization with application to communication and queueing systems,'' \emph{Synth. Lect. Commun. Netw.}, vol. 3, no. 1, pp. 1--211, Jan. 2010.

\bibitem{meta}
C. Finn, P. Abbeel, and S. Levine, ``Model-agnostic meta-learning for fast adaptation of deep networks,'' in \emph{Proc. Int. Conf. Mach. Learn. (ICML)}, 2017, pp. 1126--1135.

\bibitem{Rahmani2022}
M. Rahmani, M. Bashar, M. J. Dehghani, P. Xiao, R. Tafazolli, and M. Debbah, ``Deep reinforcement learning-based power allocation in uplink cell-free massive {MIMO},'' in \emph{Proc. IEEE Wireless Commun. Netw. Conf. (WCNC)}, 2022, pp. 459--464.

\bibitem{Secure2022Lu}
W. Lu, Y. Ding, Y. Gao, Y. Chen, N. Zhao, Z. Ding, and A. Nallanathan, ``Secure {NOMA}-based {UAV}-{MEC} network towards a flying eavesdropper,'' \emph{IEEE Trans. Commun.}, vol. 70, no. 5, pp. 3364--3376, Mar. 2022.

\bibitem{3D2019You}
C. You and R. Zhang, ``{3D} trajectory optimization in {Rician} fading for {UAV}-enabled data harvesting,'' \emph{IEEE Trans. Wireless Commun.}, vol. 18, no. 6, pp. 3192--3207, Apr. 2019.

\bibitem{powell2007approximate}
W. B. Powell, \emph{Approximate Dynamic Programming: Solving the Curses of Dimensionality}. Hoboken, NJ, USA: John Wiley \& Sons, 2007.

\end{thebibliography}

\end{document}